\documentclass
[10pt, oncecolumn,superscriptaddress,secnumarabic,amssymb,amsmath,nobibnotes,aps,prd,showpacs,nofootinbib]{revtex4}%
\usepackage{graphicx}
\usepackage{epsf}
\usepackage{bm}
\usepackage{amsmath}
\usepackage{amsfonts}
\usepackage{amssymb}
\usepackage{epstopdf}
\usepackage{natbib}
\usepackage{rotating}
\usepackage{pdflscape}
\usepackage{adjustbox}
\usepackage{color}
\usepackage{dcolumn}
\usepackage{ulem}
\usepackage{times}
\usepackage{appendix}
\usepackage{enumitem}
\usepackage{booktabs}
\usepackage{multirow}
\usepackage{orcidlink}

\begin{document}
\title{The Possibility of a Non-Lagrangian Theory of Gravity}

\author{Celia Escamilla-Rivera\orcidlink{0000-0002-8929-250X}}
\email{celia.escamilla@nucleares.unam.mx}
\affiliation{Instituto de Ciencias Nucleares, Universidad Nacional Aut\'onoma de M\'exico, Circuito Exterior C.U., A.P. 70-543, M\'exico D.F. 04510, M\'exico.}

\author{J\'ulio C. Fabris\orcidlink{0000-0001-8880-107X}}
\email{julio.fabris@cosmo-ufes.org}
\affiliation{N\'ucleo Cosmo-UFES \& Departamento de F\'isica, Universidade Federal do Esp\'irito Santo,
Av. Fernando Ferrari 514, Vit\'oria~29075-910, Brazil. \\
National Research Nuclear University ``MEPhI'',
Kashirskoe sh. 31, Moscow 115409, Russia.}

\pacs{04.50.Kd, 98.80.-k, 04.60.-m} 

\begin{abstract}
General Relativity resembles a very elegant crystal glass: If we touch its principles, that is, its Lagrangian, there is a risk of breaking everything. Or, if we will, it is like a short blanket: Curing some problems creates new problems. 
This paper is devoted to bring to light the reasons why we pursue the possibility of a non-Lagrangian theory of gravity 
under the hypothesis of an extension of the original general relativity with an ansatz inspired in the fundamental principles of classical and quantum physics.\end{abstract}

\maketitle


\section{Introduction}

In physics it is expected that all the fundamental laws can be written in terms of an action principle. In fact, the action principle has many advantages since the content of the theory, with its symmetries, can be more visible. 
Noether's theorem, which states that every Lie symmetry of the action corresponds to a conservation law,
can be proven directly from the Lagrangian and its corresponding action {\cite{Halder:2018kyp}}. An example of this consequence is when the Lagrangian of a particular system is invariant under infinitesimal translations in time and therefore the system total energy is conserved. The approach based on an action principle is generally considered, due to this connection of symmetries and conservation laws, as aesthetically and mathematically very appealing.

Traditionally, in gravitational physics the action $S$ is an integral
of some specific function of the fields 
taken over  
the four-dimensional spacetime and can be  constructed with the Ricci scalar $R$ and a function of the matter fields $\phi$, $\mathcal{L}_{m}(\phi,g_{\mu\nu})$. This leads to the Einstein--Hilbert action, which resulting variation gives the equations of motion for the fields. This simple idea can be articulated in Lovelock's theorem {\cite{Navarro:2010zm}}, which states: ``GR equations are the only gravitational field equations {in four dimensions} constructed solely from the metric, are local, are not more than second order in derivatives\footnote{Most higher-derivative theories are unstable according to Ostrogradsky's theorem \cite{Woodard:2015zca}.}
 and are derived from a Lagrangian''. 
Following this statement, the action can be built as a sum of subsequently higher-order curvature terms in a general number of dimensions, each of them with a coupling constant. This description has dynamical terms in the equations of motion which do not contribute to them and relate the number of dimensions to the curvature order \cite{Bueno:2016dol}.
Furthermore, according to some studies, in an infinite-order higher derivative, i.e. non-local theory of gravity, it is possible to avoid pathologies while recovering GR at low energies \cite{Biswas:2011ar}, {e.g., in \cite{Li:2015bqa} was presented a series of exact solution singularities for a class of weakly non-local theories of gravity below properties as unitarity and super-renormalizability or finiteness.}
However, the problems with singularities---either in cosmology and in black hole physics\footnote{We can add to these problems the tensions in the Standard Cosmological Model ($\Lambda$CDM), like those of $\sigma_8$ and $H_0$. \cite{DiValentino:2020zio,DiValentino:2020vvd}}---point out that GR is an incomplete theory, leading to temptations to modify it. 
The possible modifications are classified into those breaking fundamental assumptions or containing additional fields and even massive graviton(s) {\cite{Ezquiaga:2018btd}}.  We can modify the geometric part of $S$ by including additional degrees of freedom (massive gravity has three degrees of freedom) or higher powers of $R$.
Other modifications involve adding scalar, vector and tensor fields, higher dimensions, higher derivatives, minimally (or non-minimally) coupled scalar fields and, finally, theories in which the Lorentz invariance is violated.

\begin{figure}
\includegraphics[width=0.7\textwidth,origin=c,angle=0]{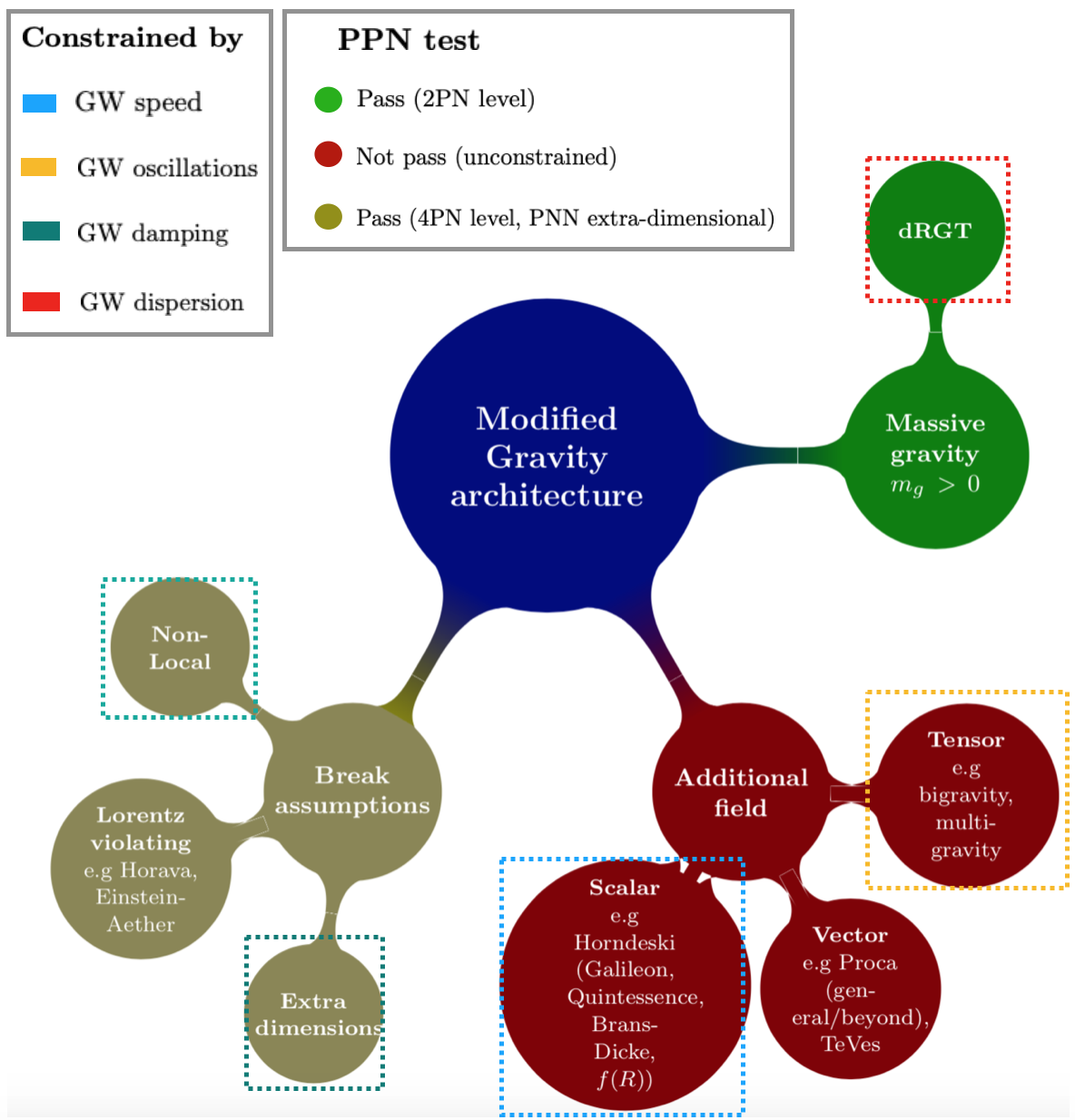}
     \caption{A modified gravity architecture, which gives a representation of some possible ways of modifying GR through breaking the Lovelock's theorem along with some examples. Theories are classified in order to avoid Lovelock’s restrictions. {In the green region we have theories as de Rham--Gabadadze--Tolley (dRGT) massive gravity theories, and cases with $m_g >0$. In the red region we have theories with additional field as scalar (Horndeski) which includes cases as Galileon, Quintessence, Brans--Dicke and $f(R)$. Moreover, Vector theories such as Proca and Tensor--Vector--Scalar (TeVes) are presented and, finally, bigravity theories. In the olive  region we have an example of theories that are constructed in the break assumption formalism as non-local ones, extra dimensions and Lorentz violating.}
     The main gravitational wave (GW) test of each theory is framed at the top, where the following was considered: GW speed (blue color) and dispersion (red color) refers to constraints from late time GW sources. GW oscillations (yellow color) refer to these oscillations during the propagation introducing a modulation of the GW amplitude.
     GW damping (green color) refers to probing the damping of GWs using standard sirens from $d^{\text{GW}}_{L}$ {(luminosity distance of GWs)} and $d^{\text{EM}}_{L}$ {(electro-magnetic luminosity distance)}. Theories that are constrained by these GW observables are enclosed by a dashed square of each GW color, respectively.
     This figure was designed based on the ideas and discussions behind the construction of theoretically sensible modified gravity theories \cite{Clifton:2011jh,Baker:2012zs,Jaime:2018ftn}. Classification of parameterised post-Newtonian (PPN) tests was based on approximations to obtain the PPN parameters at 1, 2, 4 PN level~\cite{PPN_ref,PPN_ref2,PPN_ref3}. Moreover, a PPN extension for a five-dimensional metric was included \cite{Xu:2007dc}. Notice that theories that are in agreement with this PNN test are filled in their respective colors: 2PN level (solid green color), unconstrained (solid red color) and 4PN level/PNN extra-dimensional (olive color).
     }
    \label{fig:mg_diagram}
\end{figure}

Historically, one of the first modifications of GR was the Brans--Dicke (BD) theory, which incorporates a possible variation of the cosmological coupling $G$ in a covariant way and tries to implement the Machian principle {\cite{Namavarian:2016swt}}. Unfortunately, BD quickly faces two problems: Observational constraints make this theory almost indistinguishable from GR and it has been revealed to be less {Machian} than the latter. 
More complex couplings of scalar field $\phi$ and its derivative can be implemented (see Figure \ref{fig:mg_diagram}), keeping some special features as second order differential equations. In this figure we can see a view of the spectrum of theories, deviating from GR. Together with the different ways of modifying GR, some of the most important example theories are depicted as well. The figure is drawn by thinking about breaking some of the conditions in the Lovelock's theorem. It turns out that some parts of the figure are connected. For example, some theories which add geometrical invariants can be rewritten as scalar--tensor theories. Furthermore,  theories can be part of multiple branches of such a figure.

In this case, Horndeski-type theories (HT) are very appealing since they encompass 4D Lorentz invariant actions whose metric and field variation lead to second order equations of motion. Nevertheless, enormous care must be taken, e.g., with their tests using GWs. The curvature-based gravitational formulation (namely, $f(R)$) theories modify the action of GR with an arbitrary function of $R$ that may recover the original GR at high-curvature regions and can solve the problems at cosmological scales, but at the price of being in disagreement with PPN tests.
Different possibilities will probably never reach an end, but the features of GWs seem to imply the presence of a substantial gravitational slip---predicted to be very small in GR---which can be used to rule out gravity theories.

The paradigm of constructing a new physics theory from an action principle may, however, lead to insurmountable restrictions. 
Perhaps the 
persistencies of the singularity problems in GR and its extension together with the intriguing dark sector in cosmology may
indicate the need for a more radical departure from at least some of the usual approaches to constructing a physical theory.
From a historical point of view, it was not until later that the role of diffeomorphism (in the case of GR) and gauge transformations (in the case of Maxwell's electromagnetism theory) were fully clarified. At such a time, Maxwell created an invariant Lorentz theory and a genuine field theory before special relativity was formulated and modern theories of fundamental interactions were established; hence, the implications of a symmetry profile were invoked.

In the light of this fact, in this paper we present a reason and a possible solution to the question: Is it possible that there are viable theories with identically conserved field equations 
that are not derivable from a variational principle?
It is also expected that symmetries must be contained into a Lagrangian to cope with the quantum aspects of nature.
For example, the quantum field theory provides an incredibly successful description of all known non-gravitational phenomena, with agreement between predictions and experiment \cite{Baryshev:2020fps}; moreover, if these quantum effects in GR cannot be quantified, this must undermine our satisfaction with the experimental success of its classical predictions. As an extension, recently, strong efforts have been made towards understanding the quantum gravity problem addressed inside theories of gravity, which can be tested by space-based experiments like gyroscopes in general relativity (GINGER) \cite{Capozziello:2021goa}. 

However, we must remember that
quantum effects lead to an effective theory that cannot always be reproduced classically through an action respecting the Lovelock theorem. Sometimes an action can be identified, but with higher derivative terms. Moreover, the results depend strongly on the matter content. Geometric terms are inevitable in the so-called renormalizable stress-energy tensor\footnote{See the result of a renormalizable stress-energy tensor in Sec. 6.2 of Ref.~\cite{Birrell}.}

\section{A non-Lagrangian Theory of Gravity}

What is lost by abandoning an action formalism?
We can identify a given geometric (or field, in the case of electromagnetism) structure that is the base of a new theory. 
However, what do couples to this structure matter? We know quite well how a given field enters into the equations in flat space. In a curved space, geometry can be coupled by keeping the simplest action structure.
 One example can be given by a theory (without action):
\begin{eqnarray}
R_{\mu\nu} - \frac{1}{2}g_{\mu\nu}R - g_{\mu\nu}\Lambda = 8\pi G\biggr\{1 + \alpha RT\biggl\} T_{\mu\nu},
\end{eqnarray}
with $\alpha$ being a constant. This kind of theory belongs to a structure in which the geometric part comes from the Einstein--Hilbert action,
but not the matter sector. In a vacuum everything is straightforward, but not in the presence of matter. The consequences become relevant mainly in the strong field regime. 

A successful option beyond GR can be to ignore 
the Lagrangian formalism and to implement the variations of G and $\Lambda$ directly in the field equations, with
these quantities as functions of the ambient\footnote{This has some similarities with the screening mechanism of the $f(R)$ and Hordenski theories, but the spirit is very different.}. At the same time, we  proceed a step further in trying to implement the Machian principle, in the sense that the interaction terms depend on the curvature and density, quantities that are anyway connected through the gravitational equations. 
In this conception, the PPN parameters are essentially preserved, 
at least at lower orders, because the new terms contribute at high orders\footnote{See reference \cite{Toniato:2019rrd} where the PPN parameters have been investigated for the Palatini gravity case, which contains some formal similarities with our proposal here.}.

According to later ideas, to design a {non-Lagrangian theory of gravity} we require that in a vacuum the field equations can be reduced to the standard ones, but in the presence of matter terms we should observe notable departures. These departures will be evidence that standard conservation laws do not always hold in a final theory of nature, mathematically speaking: $\nabla^{\nu} T_{\mu\nu}\neq 0$, an extended notion of the conservation laws may be required.
Our proposal has clear connections with the running of fundamental constants due to quantum effects, but implemented in a covariant way, like a {natural} attempt to incorporate these effects that are inevitably important in strong field scenarios\footnote{Some e.g in the literature are Rastall's gravitational theory and $f(R,T)$ theories \cite{Batista:2011nu}.}.
Therefore, we set a {natural generalisation} by considering two scenarios 
in a flat Friedman--Lemaître--Robertson--Walker (FLRW) filled by a perfect fluid with density $\rho$ as the matter source: 

\begin{itemize}
\item Varying G as $G (R, T)$
  \begin{eqnarray}\label{eq:vgc}
R_{\mu\nu} -\frac{1}{2}Rg_{\mu\nu} -\Lambda g_{\mu\nu} =G(R,T) T_{\mu\nu}, \\
H^{2}_{\kappa(R,T)} =\frac{1}{3}\left(G(R,T)\rho(t) +\Lambda\right), \quad\quad \\ 
\nabla^\nu T_{\mu\nu} = -\frac{\nabla^\nu G(R,T)}{G (R,T)} T_{\mu\nu}.
 \end{eqnarray}
 
\item{Varying $\Lambda$ as $\Lambda (R,T)$} 
  \begin{eqnarray}\label{eq:vcc}
R_{\mu\nu} -\frac{1}{2}Rg_{\mu\nu} -\Lambda(R,T) g_{\mu\nu} =\kappa T_{\mu\nu},\\ 
 H^{2}_{\Lambda(T)} =\frac{1}{3}\left(\kappa\rho(t) +\Lambda(R,T)\right), \\ 
 \nabla^\nu T_{\mu\nu} = -\frac{1}{\kappa} \nabla^\nu \Lambda(R,T) T_{\mu\nu}. 
 \end{eqnarray}
 
 \end{itemize}
 
\nointerlineskip

\section{Possible Solutions}

Let us consider 
\begin{eqnarray}
R_{\mu\nu} - \frac{1}{2}g_{\mu\nu}R = G(R,T)T_{\mu\nu},
\end{eqnarray}
redefined as
\begin{eqnarray}
G(R,T)T_{\mu\nu} = G_0\tilde T_{\mu\nu}, \label{eq:einsteinsred}
\end{eqnarray}
where $G_0$ is a constant. We can consider a fluid energy-momentum tensor
\begin{eqnarray}
T_{\mu\nu} = (\rho + p)u_\mu u_\nu - pg_{\mu\nu}.
\end{eqnarray}

With these assumptions, Equation (\ref{eq:einsteinsred}) becomes
\begin{eqnarray}
R_{\mu\nu} - \frac{1}{2}g_{\mu\nu}R = G_0\tilde T_{\mu\nu},  \quad \text{and}\quad
{\tilde T^{\mu\nu}}_{;\mu} = 0.
\end{eqnarray}

Notice that the content is almost essentially the same as GR, with some subtleties. 
This redefinition leads to
\begin{eqnarray}
\tilde\rho = G(R,T)\rho,\\
\tilde p = G(R,T)p.
\end{eqnarray}

Everything seems to be trivial, so far. A more interesting example is given by an interacting model corresponding to a modified $\Lambda$CDM:
 \begin{eqnarray}
 T^{\mu\nu} = T^{\mu\nu}_m + T^{\mu\nu}_\Lambda.
 \end{eqnarray}
 where the subscripts stand for {matter} and {cosmological term}, with
 \begin{eqnarray}
 p_m = 0 \quad , \quad p_\Lambda = - \rho_\Lambda.
 \end{eqnarray}
 
 We choose now
 \begin{eqnarray}
 G(R,T) = G_0(1 + \alpha T_m) = G_0(1 + \alpha \rho_m).
 \end{eqnarray}
 
To assure the structure formation we impose,
 \begin{eqnarray}
 {T^{\mu\nu}_m}_{;\mu} = 0.
 \end{eqnarray}
 
With this hypothesis, the Bianchi identities imply
 \begin{eqnarray}
 G_{;\mu}(T^{\mu\nu}_m + T^{\mu\nu}_\Lambda) + G{T^{\mu\nu}_\Lambda}_{;\mu} = 0.
 \end{eqnarray}
 
 In a flat cosmological background these equations can be written as
 \begin{eqnarray}
 \rho_m = \frac{\rho_{m0}}{a^3},\\
 \rho_{\Lambda} = \frac{\rho_{\Lambda0} - \frac{\alpha}{2}\rho_m^2}{1 + \alpha\rho_m},
 \end{eqnarray}
 where $\rho_{m0}$ and {$\rho_{\Lambda0}$ are constants}. When $\alpha \rightarrow 0$, $\rho_\Lambda = \rho_{\Lambda0}$, $\Lambda$CDM is recovered. In~particular, observational tests such as those of supernovae type Ia are fitted as in the standard model. However,  
if $\alpha \neq 0$ there are strong deviations from the standard model, which must be more significant in the strong regime. If $\alpha$ is close to zero, the deviations will be small.

The Friedmann equation becomes,
{
\begin{eqnarray}
\biggr(\frac{\dot a}{a}\biggl)^2 = \frac{8\pi G_0}{3}\biggr\{\rho_m\biggr(1 + \frac{\alpha}{2}\rho_m  \biggl) + \rho_{\Lambda0}\biggl\}.
\end{eqnarray}
}

For $\alpha=0$, the $\Lambda$CDM model is recovered. However, for $\alpha \gg 1$, we have,
\begin{eqnarray}
\label{gg}
H^2= \biggr(\frac{\dot a}{a}\biggl)^2 = \frac{8\pi G_0\alpha}{6}\rho_m^2.
\end{eqnarray}

For very large scale factor values the matter term disappears and we have essentially the $\Lambda$CDM model, whereas the limit $\alpha \gg 1$ may play a more relevant role in the primordial universe because in this epoch the matter term dominates.
The solution in this high density regime is $a \propto t^{1/3}$, typical of stiff matter; the expansion of the universe is slower than predicted by the standard model, even if matter is pressureless. We remark that $\alpha$ must be positive.
A straightforward calculation leads to the following equation for the matter density contrast $\delta_m$:
\begin{eqnarray}
\ddot\delta_m + 2 H\dot\delta_m - 4\pi G_0(\rho_m + 4\alpha\rho_m^2)\delta_m = 0.
\end{eqnarray}

Again, we find the usual perturbative behaviour of the $\Lambda$CDM for the late universe, since when $\alpha$ is very small the latter equation becomes
\begin{eqnarray}
\ddot\delta_m + 2 H\dot\delta_m - 4\pi G_0\rho_m\delta_m = 0.
\end{eqnarray}

For the early universe, {after using (\ref{gg}) in the last term in the left-hand-side 
and the expression for the scale factor at that regime,} we obtain,
\begin{eqnarray}
\ddot\delta_m + 2 H\dot\delta_m - 16\pi G_0\alpha \rho_m^2\delta_m = 0,
\end{eqnarray}
where the growing mode behaves in this limit as
\begin{eqnarray}
\delta_m^+ \propto t^{4/3}.
\end{eqnarray}

The perturbations grow much faster than in the $\Lambda$CDM model during the early universe, for which $\delta_m \propto t^{2/3}$. This is an interesting feature implying that the role of dark matter to enhance the perturbations during the radiation dominated era would be less important than in the standard model.
With the perturbation growing as $t^{4/3}$, and starting from fluctuations of the order of $10^{- 5}$ (e.g., CMB), we have today that the fluctuations are of the order of $10^{3}$ with respect to the average density, which is the typical density value  for galaxies. Of course, we have made an extrapolation of the linear result to the non-linear regime. However, this extrapolation gives an idea of the new effects due to the unusual matter coupling.
In the $\Lambda$CDM model, under the same circumstances, today we have $10^{-1}$; therefore, not even the non-linear regime has been reached (hence, the need for dark matter). 
These theoretical differences in the growth of structure may offer the possibility to distinguish between modified gravity theories and the $\Lambda$CDM model. Moreover, an issue to applying current and forthcoming large scale galaxy surveys to these differences is the unavailability of numerical simulations of non-linear growth in modified theories of gravity. Some efforts in this direction, considering modified dark energy theories, have been made in order to predict a non-linear growth in such theories if the linear power spectrum fits with scale-independent or dependent modifications \cite{Laszlo:2007td}.

\section{Conclusions}

In this paper we shed some light on the idea of designing possible viable theories with identically conserved field equations 
that are not derivable from a variational principle. By setting a natural generalisation through two scenarios 
in a flat FLRW filled by a perfect fluid, we found that the perturbations grow much faster than in the standard cosmological model. This result shows that 
the role of dark matter could be less important in our approach in comparison to $\Lambda$CDM, which could bring differences in the growth of structure.

We know that GR will not be the full answer to everything, as there are questions we can ask that it is incapable of addressing. We showed that it is reasonable to explore different ways to {modify} GR, to work out the consequences, and to look for deviations. Such a breakdown of some, in principle, well established principles may lead to new approaches, which can open up unexpected perspectives. The problems that lead to the proposal of a dark sector in the standard cosmological model seem to require such a drastic conceptual jump.


\section*{Acknowledgments}
CE-R acknowledges the Royal Astronomical Society as FRAS 10147 and networking support by the COST Action CA18108. J.C.F. thanks support from FAPES (Brazil) and CNPq (Brazil). This research was funded by DGAPA-PAPIIT-UNAM through grant No. IA100220.
The authors thank the reviewers for their insightful constructive criticisms of the results.



\end{document}